\documentclass[runningheads]{llncs}

\usepackage[T1]{fontenc}
\def\doi#1{\href{https://doi.org/\detokenize{#1}}{\url{https://doi.org/\detokenize{#1}}}}
\usepackage{graphicx}
\usepackage{subfigure}

\usepackage{multicol}
\usepackage{multirow}

\usepackage{amsmath, bm}
\usepackage{mathrsfs}
\usepackage{amsfonts}
\usepackage{tablefootnote}
\usepackage{booktabs}
\usepackage{color}
\usepackage[misc]{ifsym}

\usepackage[colorlinks,
linkcolor=blue,
anchorcolor=blue,
citecolor=blue,
urlcolor=blue]{hyperref}

\usepackage{listings}
\begin{document}

%
\title{Dual-Distribution Discrepancy for Anomaly Detection in Chest X-Rays}
%
%
\author{Yu Cai\inst{1,2(\textrm{\Letter})}  \and  
Hao Chen\inst{3}  \and  
Xin Yang\inst{2}  \and  
Yu Zhou\inst{2}   \and  
Kwang-Ting Cheng\inst{1, 3}  
}

\authorrunning{Y. Cai et al.}
%
\institute{
Department of Electronic and Computer Engineering, The Hong Kong University of Science and Technology, Hong Kong, China\\ 
\email{yucai200009@gmail.com} \\ \and
School of Electronic Information and Communications, Huazhong University of Science and Technology, Wuhan, China\\ \and
Department of Computer Science and Engineering, The Hong Kong University of Science and Technology, Hong Kong, China\\
}

\maketitle              

\begin{abstract}
Chest X-ray (CXR) is the most typical radiological exam for diagnosis of various diseases. Due to the expensive and time-consuming annotations, detecting anomalies in CXRs in an unsupervised fashion is very promising. However, almost all of the existing methods consider anomaly detection as a one-class classification (OCC) problem. They model the distribution of only known normal images during training and identify the samples not conforming to normal profile as anomalies in the testing phase. A large number of unlabeled images containing anomalies are thus ignored in the training phase, although they are easy to obtain in clinical practice. In this paper, we propose a novel strategy, Dual-distribution Discrepancy for Anomaly Detection (DDAD), utilizing both known normal images and unlabeled images. The proposed method consists of two modules. During training, one module takes both known normal and unlabeled images as inputs, capturing anomalous features from unlabeled images in some way, while the other one models the distribution of only known normal images. Subsequently, inter-discrepancy between the two modules, and intra-discrepancy inside the module that is trained on only normal images are designed as anomaly scores to indicate anomalies. Experiments on three CXR datasets demonstrate that the proposed DDAD achieves consistent, significant gains and outperforms state-of-the-art methods. Code is available at \url{https://github.com/caiyu6666/DDAD}.

\end{abstract}

\section{Introduction}
Thanks to the cost-effectiveness and low radiation dose, combined with a reasonable sensitivity to a wide variety of pathologies, chest X-ray (CXR) is the most commonly performed radiological exam \cite{ccalli2021deep}. To alleviate radiologists’ reading burden and improve diagnosis efficiency, automatic CXR analysis using deep learning is becoming popular \cite{luo2021oxnet,luo2020deep}. However, annotations of CXRs are difficult to obtain because of the expertise requirements and time-consuming reading, which motivates the development of anomaly detection that requires few or even no annotations.

\begin{figure}[!t]
\centering
\includegraphics[width=\textwidth]{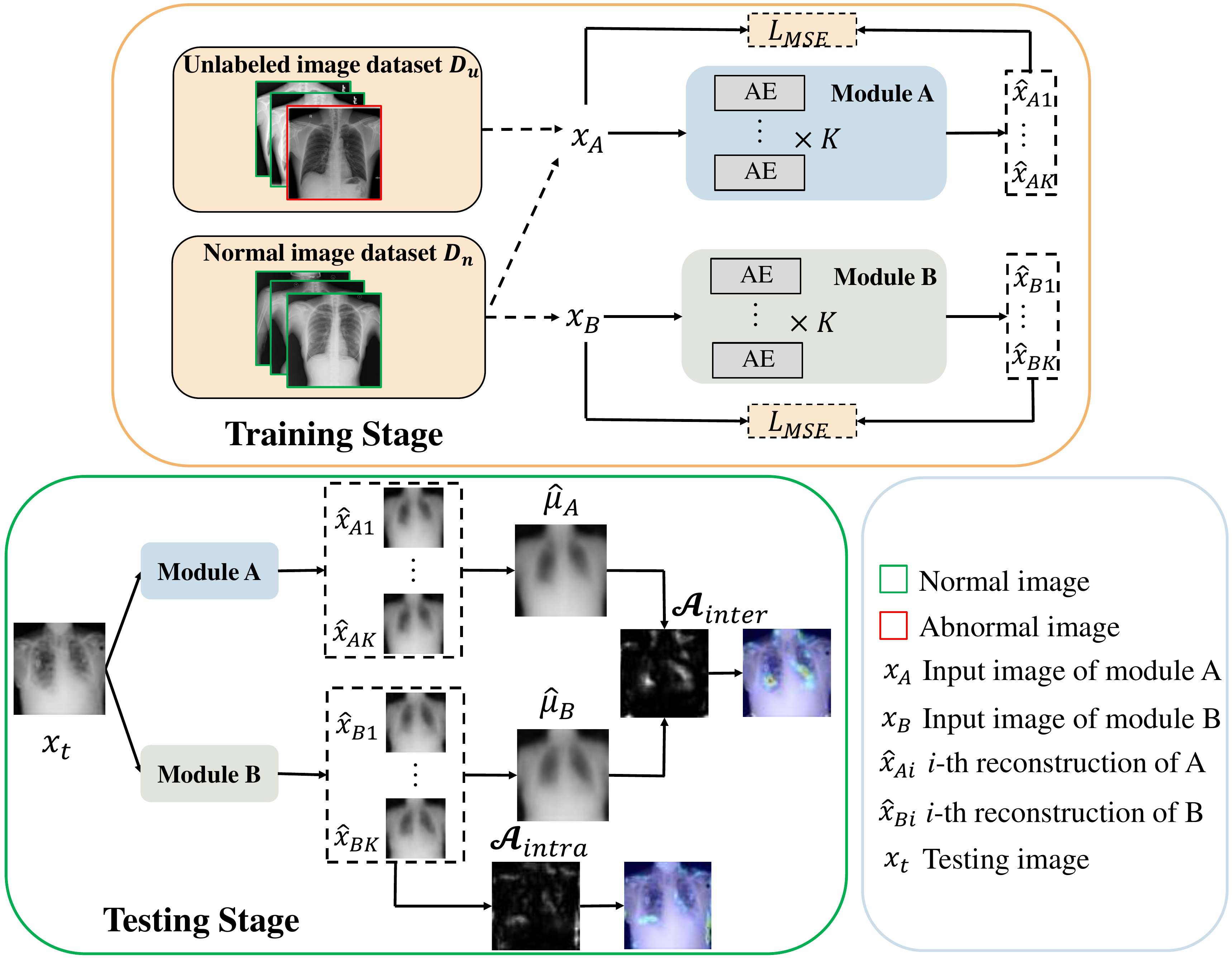}
\caption{Illustration of the proposed DDAD. To simplify the visualization, we utilize the AE as the backbone in both modules A and B here.} \label{fig1}
\end{figure}

Previously, most existing methods consider anomaly detection as a one-class classification (OCC) problem \cite{ruff2018deep}, where only normal images are utilized for training and samples not conforming to normal profiles are identified as anomalies in the testing phase. Under this setting, extensive methods based on reconstruction \cite{baur2021autoencoders} or self-supervised learning \cite{jing2020self} have been proposed for anomaly detection. Reconstruction-based methods \cite{gong2019memorizing,schlegl2019f,mao2020abnormality,akcay2018ganomaly,baur2020scale} have been proven effective. They train the reconstruction networks like variants of autoencoder (AE) to minimize the reconstruction error on normal images, while unseen abnormal images are assumed not able to be reconstructed, and in turn yield larger reconstruction errors. To avoid the reconstruction of anomaly and reduce miss detection, some methods \cite{chen2021normative,marimont2021anomaly} utilize Variational AE (VAE) \cite{kingma2013auto} to approximate the normative distribution and perform image restoration iteratively to ensure the output is anomaly-free, yielding higher difference with the abnormal input. However, the iterative restoration process is computationally complex and time-consuming. Recently, some self-supervised methods \cite{tan2020detecting,tan2021detecting,li2021cutpaste,zavrtanik2021draem} try to synthesize defects manually for training models to detect irregularities. Some \cite{tian2021constrained,sohn2020learning} also design contrastive learning frameworks to learn more effective representations. But the performance is limited by the lack of real abnormal samples. 

In summary, existing methods either use only normal images or use synthetic abnormal images on top of them during training, whose discriminative capability is limited by the lack of training on real abnormal images. Meanwhile, there are plenty of readily available unlabeled images with a reasonable anomaly rate (AR) in clinical practice, which are ignored by existing works. Based on this observation, we raise a problem: whether unlabeled images can provide effective information of abnormalities as a complement of normal images to improve the performance of anomaly detection? Motivated by this, we propose the Dual-distribution Discrepancy for Anomaly Detection (DDAD) to utilize both known normal images and unlabeled images. We demonstrate that the use of unlabeled images can significantly improve the performance of anomaly detection. The proposed DDAD consists of two modules, denoted as A and B, each of which is an ensemble of several reconstruction networks with the same architecture. During training, module A takes both known normal and unlabeled images as inputs, capturing anomalous features from unlabeled images in some way, while module B models the distribution of only known normal images. Intuitively, high discrepancy will derive between modules A and B in abnormal regions, thus inter-discrepancy is applied as an anomaly score. Besides, as module B is trained on only normal images, the reconstructions' variance will be high in abnormal regions, thus intra-discrepancy inside B can be used as another anomaly score. To the best of our knowledge, it is the first time that unlabeled images are utilized for anomaly detection. Experiments show that our method achieves significant improvement and obtains state-of-the-art performance on three CXR datasets.

\section{Method}
In this section, we introduce the proposed DDAD. Previously, most of the existing works formulate the anomaly detection as an OCC problem. That is, given a normal image dataset $D_n=\{\bm{x}_{ni}, i=1,...,N\}$ with \emph{N} normal images, and a test dataset $D_t=\{(\bm{x}_{ti}, y_i), i=1,...,T\}$ with \emph{T} annotated normal or abnormal images, where $y_i \in \{0, 1\}$ is the image label (0 for normal image and 1 for abnormal image), the goal is to train a model based on the normal image set $D_n$ which can identify anomalies in the test dataset $D_t$ during inference. Different from previous works, our proposed DDAD makes full use of unlabeled images in clinical practice. Specifically, except for the normal image dataset $D_n$, we also utilize a readily available unlabeled image dataset $D_u=\{\bm{x}_{ui}, i=1,...,M\}$ with \emph{M} unlabeled images including both normal and abnormal images, to improve the performance of anomaly detection. 

\subsection{Dual-distribution Modeling}
In the proposed DDAD as shown in Fig.~\ref{fig1}, we use two modules, denoted as A and B, trained on different datasets to model the dual-distribution. Each module is an ensemble of $K$ reconstruction networks with the same architecture but different random initialization of parameters and random shuffling of training samples, trained by the Mean Squared Error (\emph{MSE}) Loss to minimize reconstruction errors on the training set. Specifically, module A is trained on both normal image dataset $D_n$ and unlabeled image dataset $D_u$ as:
\begin{equation}
\mathcal{L}_A = \frac{1}{N+M} \sum_{\bm{x}_A \in D_n \cup D_u} \sum_{i=1}^{K} \Vert\bm{x}_A-\bm{\hat{x}}_{Ai}\Vert ^2,
\label{eq1}
\end{equation}
\noindent
where \emph{N} and \emph{M} are sizes of datasets $D_n$ and $D_u$ respectively, $\bm{x}_A$ is the input image of module A, and  $\bm{\hat{x}}_{Ai}$ is the reconstruction of $\bm{x}_A$ from $i$-th network in module A.
Similarly, the loss function of module B trained on only normal image dataset $D_n$ can be written as:
\begin{equation}
\mathcal{L}_B = \frac{1}{N} \sum_{\bm{x}_B \in D_n} \sum_{i=1}^{K} \Vert\bm{x}_B-\bm{\hat{x}}_{Bi}\Vert ^2.
\label{eq2}
\end{equation}

Through this way, module A captures effective information of abnormalities from the unlabeled dataset as a complement of normal images. Therefore, high discrepancy between A and B will derive at abnormal regions as module B never sees anomlies during training. Besides, based on the theory of Deep Ensemble \cite{lakshminarayanan2017simple}, networks in module B will also show high uncertainty (i.e., intra-discrepancy) on unseen anomlies. Our anomaly scores (described in Sec.~\ref{Anomaly Score}) are designed based on the above analysis to indicate abnormal regions subsequently.

\subsection{Dual-distribution Discrepancy-based Anomaly Scores} \label{Anomaly Score}
Given a testing image $\bm{x}_t$, the pixel-wise reconstruction error $\mathcal{A}_{rec}^p = (\bm{x}_t^p - \bm{\hat{x}}_t^p)^2$ has been widely used as anomaly score previously. Base on the proposed DDAD and above analysis, we propose to use intra- and inter-discrepancy to indicate anomalies as following:
\begin{equation}
\mathcal{A}_{intra}^p = \sqrt{\frac{1}{K} \sum_{i=1}^{K} (\hat{\mu}_B^p - \hat{x}_{Bi}^p)^2};~~~\mathcal{A}_{inter}^p = \vert \hat{\mu}_A^p - \hat{\mu}_B^p \vert.
\end{equation}
Here $p$ is the index of pixels, $\bm{\hat{\mu}}_A$ and $\bm{\hat{\mu}}_B$ are average maps of reconstructions from modules A and B, respectively. As shown in Fig.~\ref{fig1}, our discrepancy maps can indicate potential abnormal regions based on the pixel-wise anomaly scores. The image-level anomaly score is obtained by averaging the pixel-level scores in each image.

Compared with $\mathcal{A}_{rec}$, our anomaly scores for the first time consider the discrepancy between different distributions, leading to stronger discriminative capability. Intuitively, higher AR in unlabeled dataset will lead to greater difference between two distributions on abnormal regions, deriving more competitive $\mathcal{A}_{inter}$. Fortunately, experiments in Fig.~\ref{fig2} demonstrate that even if AR is 0 we can still achieve a consistent improvement compared with the reconstruction baseline, while a low AR can lead to significant boost. Besides, our discrepancies are all computed among reconstructions, rather than between the input and reconstruction as $\mathcal{A}_{rec}$ does. This can reduce the false positive detection caused by reconstruction ambiguity of AE around high frequency regions \cite{baur2021autoencoders,mao2020abnormality}.

\subsection{Uncertainty-refined Dual-distribution Discrepancy}
Due to the reconstruction ambiguity of AE, high reconstruction errors often appear at high frequency regions, e.g., around normal region boundaries, leading to false positive detection. To address this problem, AE-U \cite{mao2020abnormality} proposed to refine the $\mathcal{A}_{rec}$ using estimated pixel-wise uncertainty. It generates the reconstruction $\hat{\bm{x}}_i$ and corresponding uncertainty $\bm{\sigma}^2(\bm{x}_i)$ for each input $\bm{x}_i$, trained by:

\begin{equation}
\mathcal{L} = \frac{1}{NP} \sum_{i=1}^{N} \sum_{p=1}^{P} \{ \frac{(x_i^p-\hat{x}_i^p)^2}{\sigma_p^2(\bm{x}_i)} + {\rm log} \sigma_p^2(\bm{x}_i) \}
\end{equation}
Training on normal images, the numerator of the first term is an \emph{MSE} loss to minimize the reconstruction error, while the $\sigma_p^2(\bm{x}_i)$ at the denominator will be learned automatically to be large at pixels with high reconstruction errors to minimize the first term. Besides, the second term drives the predicted uncertainty to be small at other regions. The two loss terms together ensures that the predicted uncertainty will be larger at only normal regions with high reconstruction errors, thus it can be used to refine the anomaly score at pixel-level.

In this work, we design a strategy similar to AE-U while adapting to DDAD well. We use AE-U as the backbone of DDAD, and utilize the uncertainty predicted by our module B, which is trained on only normal image dataset, to refine our intra- and inter-discrepancy at $p$-th pixel as:

\begin{equation}
\mathcal{A}_{intra}^p = \frac{\sqrt{\frac{1}{K} \sum_{i=1}^{K} (\hat{\mu}_B^p - \hat{x}_{Bi}^p) ^2}}{\sigma_p};~~~ \mathcal{A}_{inter}^p = \frac{\vert \hat{\mu}_A^p - \hat{\mu}_B^p \vert}{\sigma_p}.
\end{equation}
Here $\sigma_p$ is the average uncertainty predicted by AE-Us in module B.

\section{Experiments}
\subsection{Datasets}
We conduct extensive experiments on three CXR datasets: 1) RSNA Pneumonia Detection Challenge dataset\footnote[1]{\url{https://www.kaggle.com/c/rsna-pneumonia-detection-challenge}}, 2) VinBigData Chest X-ray Abnormalities Detection dataset\footnote[2]{\url{https://www.kaggle.com/c/vinbigdata-chest-xray-abnormalities-detection}}, 3) Chest X-ray Anomaly Detection (CXAD) dataset. The performance is assessed with area under the ROC curve (AUC). \textbf{RSNA dataset} contains 8,851 normal and 6,012 lung opacity images. In experiments, we use 3,851 normal images as the normal dataset $D_n$, 4,000 images with different ARs as the unlabeled dataset $D_u$, and 1,000 normal and 1,000 lung opacity images as the test dataset $D_t$. \textbf{VinBigData dataset} contains 10,606 normal and 4,394 abnormal images that include 14 categories of anomalies in total. In experiments, we use 4,000 normal images as $D_n$, 4,000 images as $D_u$, and 1000 normal and 1000 abnormal images as $D_t$. \textbf{CXAD dataset}, collected by us for this study, contains 3,299 normal and 1,701 abnormal images that include 18 categories of anomalies in total. In experiments, we use 2,000 normal images as $D_n$, 2,000 images as $D_u$, and 499 normal and 501 abnormal images as $D_t$.

\subsection{Implementation Details}
The AE in our experiments contains an encoder and a decoder. The encoder contains 4 convolutional layers with kernel size 4 and stride 2, whose channel sizes are 16-32-64-64. The decoder contains 4 deconvolutional layers with the same kernel size and stride as the encoder, and the channel sizes are 64-32-16-1. The encoder and deocder are connected by 3 fully connected layers. All layers except the ouput layer are followed by batch normalization and ReLU. For fair comparison, MemAE \cite{gong2019memorizing} and AE-U \cite{mao2020abnormality} in our experiments are modified based on this AE. All the input images are resized to $64 \times 64$. $K$ is set to 3. Each model is trained for 250 epochs using the Adam optimizer with a learning rate of 5e-4. All experiments were run on a single NVIDIA TITAN Xp GPU.

\subsection{Ablation Study}
\subsubsection{DDAD with different ARs.} 
\begin{figure}[ht]
\centering
\includegraphics[width=0.55\textwidth]{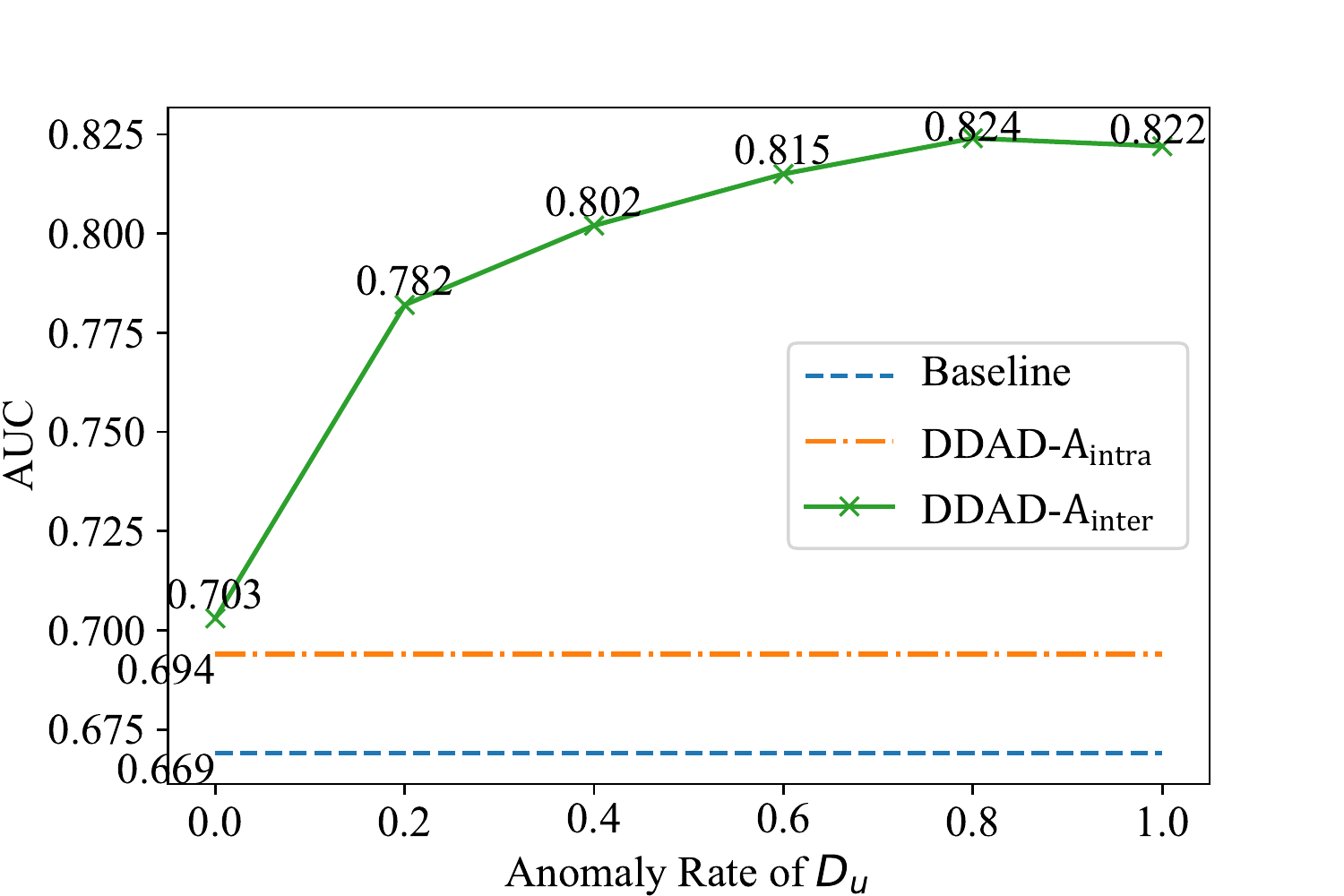}
\caption{Performance of DDAD (AE) on RSNA dataset with varying AR of $D_u$ compared with the reconstruction baseline using AE.} \label{fig2}
\end{figure}

In clinical practice, the AR of unlabeled dataset $D_u$ is unknown. In order to simulate the real scenario, we evaluate the proposed DDAD based on AE on RSNA dataset with AR of $D_u$ varying from 0 to 100\%. We use the reconstruction-based method with AE as the baseline for comparison. As shown in Fig.~\ref{fig2}, the proposed DDAD method, especially when using $\mathcal{A}_{inter}$, achieves consistent and significant improvement compared with baseline, while DDAD using $\mathcal{A}_{inter}$ performs better with the increasing AR of $D_u$. Note that $\mathcal{A}_{intra}$ is computed inside module B, thus irrelevant to AR.

More specifically, several observations from Fig.~\ref{fig2} demonstrate the superiority of our method. First, even in the extreme situation (i.e., AR is 0), the DDAD method can still achieve better performance than baseline. That's to say, we can apply the DDAD strategy in any situations and get improvement consistently regardless of AR. Intuitively, when AR is 0, dataset $D_n \cup D_u$ only contains normal images, thus module A degenerates to the same as B. However, in this situation module A is trained on a larger normal dataset than baseline, which leads to more robust models and supports the consistent improvement.
Second, even if AR is low (e.g., 20\%), the DDAD can achieve a significant improvement. That means the proposed DDAD can improve the performance considerably in clinical practice as there are always some abnormal cases.

\begin{figure}[ht]
\centering
\subfigure[Reconstruction]{
\begin{minipage}[t]{0.31\linewidth}
    \centering
	\includegraphics[width=\linewidth]{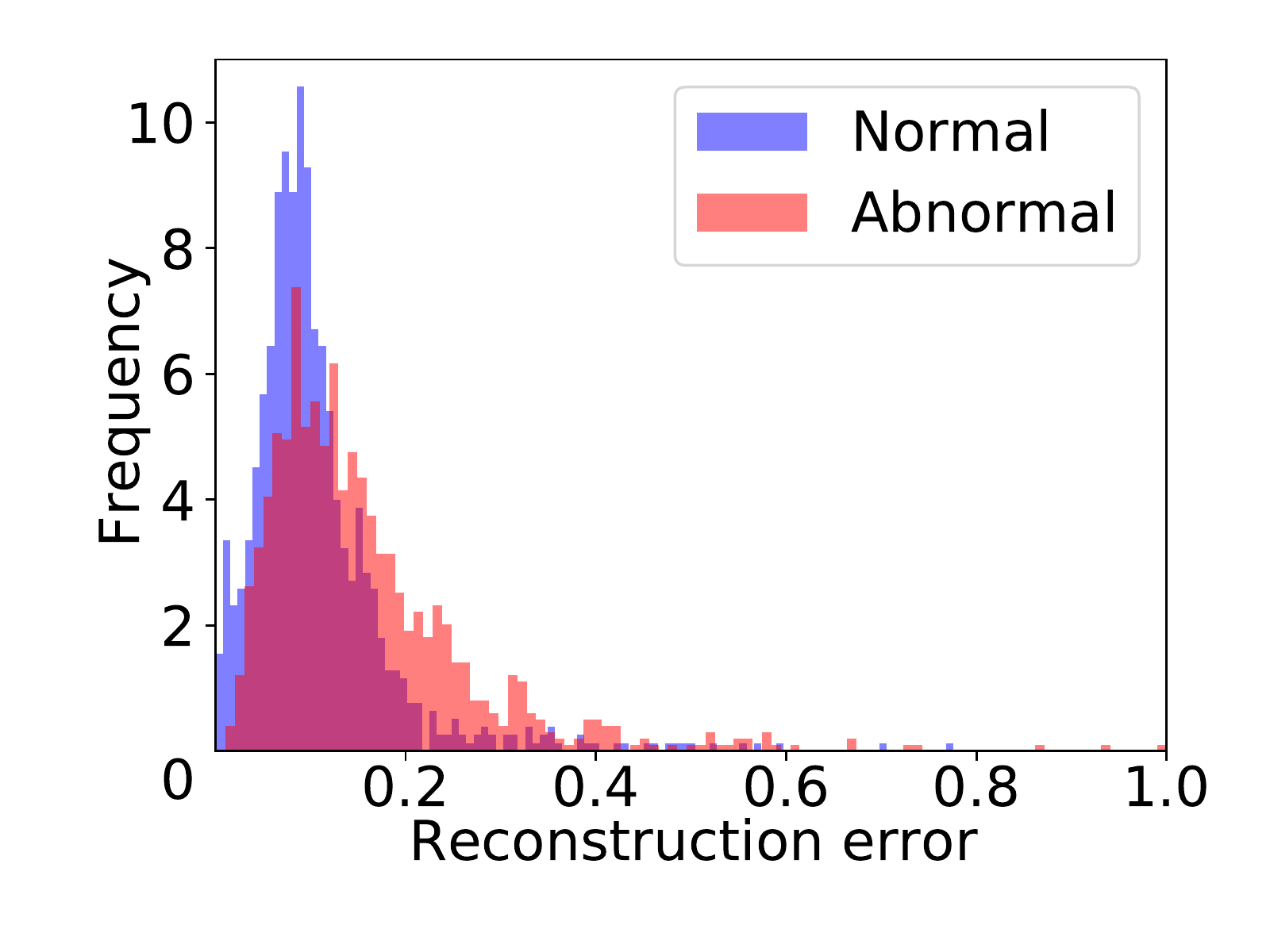}
\end{minipage}%
}
\subfigure[DDAD-$\mathcal{A}_{intra}$]{
\begin{minipage}[t]{0.31\linewidth}
	\centering
	\includegraphics[width=\linewidth]{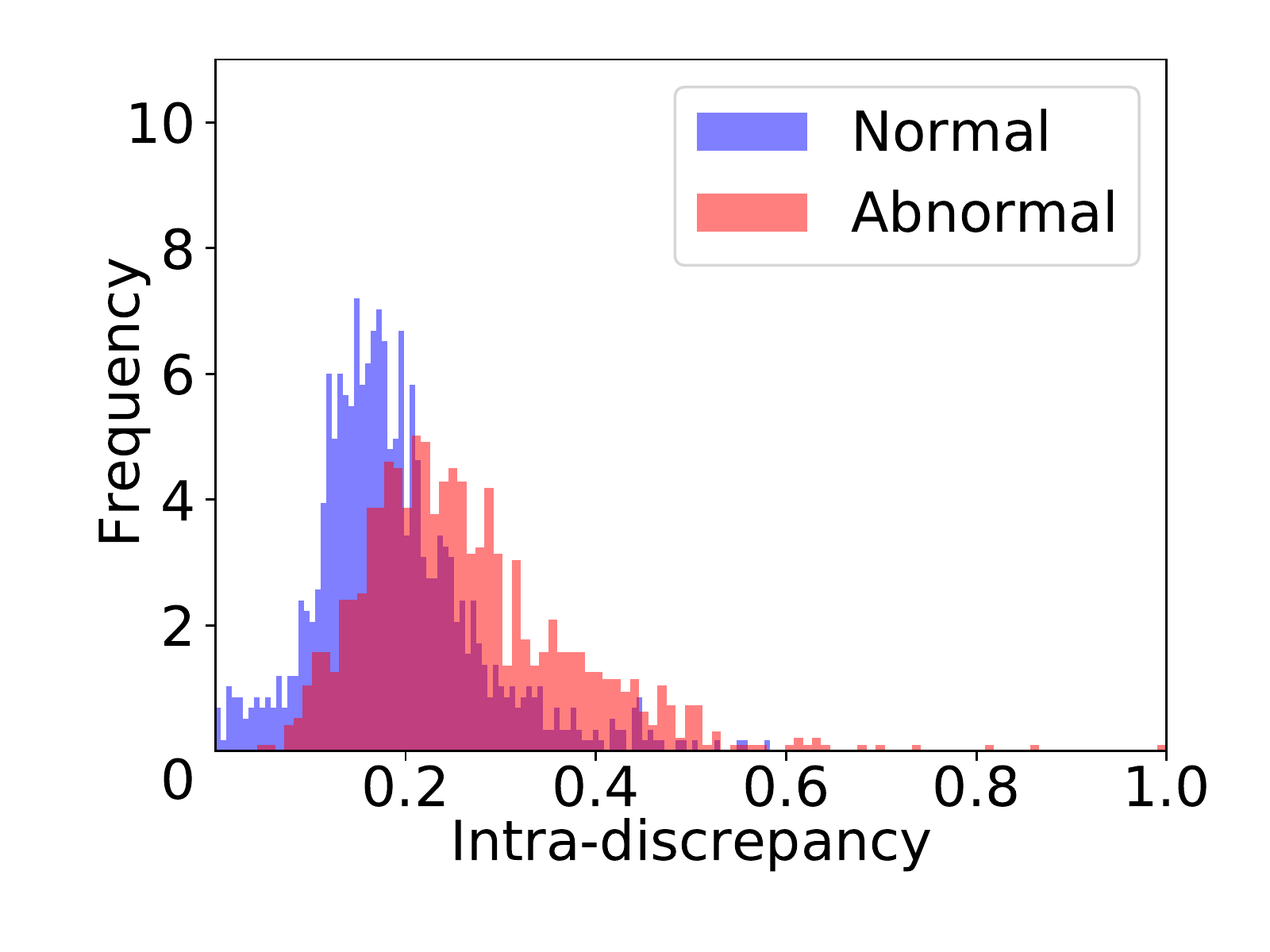}
\end{minipage}
}
\subfigure[DDAD-$\mathcal{A}_{inter}$]{
\begin{minipage}[t]{0.31\linewidth}
	\centering
	\includegraphics[width=\linewidth]{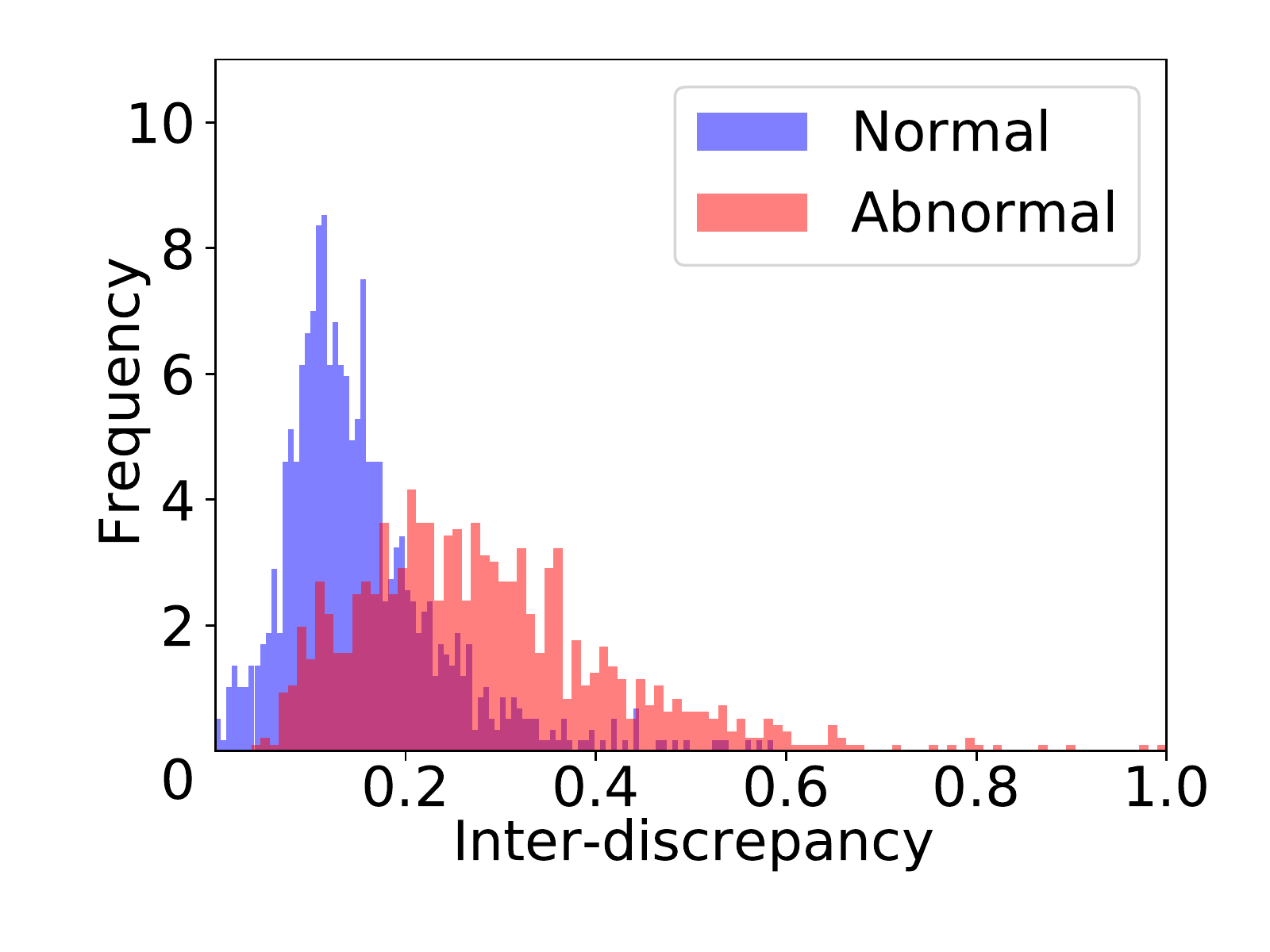}
\end{minipage}
}

\caption{Histograms of anomaly score for normal (blue) and abnormal (red) images in the test set of RSNA. The backbone is AE. Scores are normalized to [0,1].}
\label{fig3}
\end{figure}

Reference to ARs of several public large-scale CXR datasets (e.g., 71\% in RSNA and 46\% in ChestX-ray8 \cite{wang2017chestx}), we generally assume an AR of 60\% for $D_u$ in following experiments. We visualize the histograms in this setting using AE as the backbone for qualitative analysis in Fig.~\ref{fig3}. The overlap of normal and abnormal histograms in DDAD is significantly less than the reconstruction method, suggesting stronger discriminative capability for identifying anomalies.

\subsubsection{DDAD with different backbones.} Our proposed DDAD method can use any AEs' variants as the backbones. In order to further explore the advantages, DDAD built on different backbones are compared with corresponding reconstruction baselines (Rec.) in Table~\ref{tab1}. AR of $D_u$ is set to 60\% for all DDAD methods. The results show that DDAD based on AE, MemAE \cite{gong2019memorizing} and AE-U \cite{mao2020abnormality} can all outperform corresponding baselines on three CXR datasets by a large margin. In terms of AUC, DDAD-$\mathcal{A}_{inter}$ improves baselines of AE, MemAE and AE-U by 14.6\%, 10.8\% and 4.3\% on RSNA dataset, by 15.1\%, 13.2\% and 12.1\% on VinBigData dataset, by 6.5\%, 3.9\% and 5.0\% on CXAD dataset, respectively. DDAD-$\mathcal{A}_{intra}$ improves baselines of AE, MemAE and AE-U by 2.5\%, 4.9\% and 0.6\% on RSNA dataset, by 4.2\%, 3.7\% and 0.5\% on VinBigData dataset, by 4.2\%, 3.4\% and 2.8\% on CXAD dataset, respectively.

We also test the ensemble of $K$ reconstruction models using $\mathcal{A}_{rec}$, demonstrating that simple ensemble has no significant improvement.

\begin{table}[ht]
\centering
\caption{Performance of different methods built on three backbones. Bold numbers indicate the best results for each baseline.}\label{tab1}
\resizebox{\linewidth}{!}{
\begin{tabular}{c|c|ccccccccc}
\hline
\multirow{3}{*}{\textbf{Method}} &
  \multirow{3}{*}{\textbf{Anomaly Score}} &
  \multicolumn{9}{c}{\textbf{AUC}} \\ \cline{3-11}
 &
   &
  \multicolumn{3}{c|}{RSNA} &
  \multicolumn{3}{c|}{VinBigData} &
  \multicolumn{3}{c}{CXAD} \\ \cline{3-11} 
 &
   &
  AE &
  MemAE &
  \multicolumn{1}{c|}{AE-U} &
  AE &
  MemAE &
  \multicolumn{1}{c|}{AE-U} &
  AE &
  MemAE &
  AE-U \\ \hline
Rec. &
  \multirow{2}{*}{$\mathcal{A}_{rec}$} &
  0.669 &
  0.680 &
  \multicolumn{1}{c|}{0.867} &
  0.559 &
  0.558 &
  \multicolumn{1}{c|}{0.738} &
  0.556 &
  0.560 &
  0.664 \\
Rec. (ensemble) &
   &
  0.669 &
  0.670 &
  \multicolumn{1}{c|}{0.866} &
  0.555 &
  0.553 &
  \multicolumn{1}{c|}{0.731} &
  0.550 &
  0.552 &
  0.659 \\ \hline
\multirow{2}{*}{DDAD} &
  $\mathcal{A}_{intra}$ &
  0.694 &
  0.729 &
  \multicolumn{1}{c|}{0.873} &
  0.601 &
  0.595 &
  \multicolumn{1}{c|}{0.743} &
  0.598 &
  0.594 &
  0.692 \\
 &
  $\mathcal{A}_{inter}$ &
  \textbf{0.815} &
  \textbf{0.788} &
  \multicolumn{1}{c|}{\textbf{0.910}} &
  \textbf{0.710} &
  \textbf{0.690} &
  \multicolumn{1}{c|}{\textbf{0.859}} &
  \textbf{0.621} &
  \textbf{0.599} &
  \textbf{0.714} \\ \hline
\end{tabular}
}
\end{table}

\subsection{Comparison with State-of-the-art Methods}
We compare our method with four state-of-the-art (SOTA) methods, including AE, MemAE \cite{gong2019memorizing}, f-AnoGAN \cite{schlegl2019f} and AE-U \cite{mao2020abnormality}. Results in Table~\ref{tab2} show that while all DDAD methods outperform the reconstruction methods using the same backbones, the DDAD built on AE-U outperforms all other methods and achieves state-of-the-art performance.

\renewcommand{\thefootnote}{\fnsymbol{footnote}}

\begin{table}[ht]
\centering
\caption{Comparison with SOTA methods. Bold face with underline indicates the best, and bold face for the second best.}\label{tab2}
\begin{tabular}{c|c|ccc}
\hline
\multirow{2}{*}{\textbf{Method}} & \multirow{2}{*}{\textbf{Anomaly Score}} & \multicolumn{3}{c}{\textbf{AUC}} \\ \cline{3-5} 
                              &                                      & RSNA                 & VinBigData           & CXAD                 \\ \hline
AE                            & \multirow{4}{*}{$\mathcal{A}_{rec}$} & 0.669                & 0.559                & 0.556                \\
MemAE                         &                                      & 0.680                & 0.558                & 0.560                \\
f-AnoGAN\tablefootnote{Consistent with \cite{schlegl2019f}, we combine pixel-level and feature-level reconstruction errors as $\mathcal{A}_{rec}$ for f-AnoGAN.}
                              &                                      & 0.798                & \textbf{0.763}       & 0.619                \\
AE-U                          &                                      & 0.867                & 0.738                & 0.664                \\ \hline
\multirow{2}{*}{Ours (AE)}    & $\mathcal{A}_{intra}$                & 0.694                & 0.601                & 0.589                \\
                              & $\mathcal{A}_{inter}$                & 0.815                & 0.710                & 0.621                \\ \hline
\multirow{2}{*}{Ours (MemAE)} & $\mathcal{A}_{intra}$                & 0.729                & 0.595                & 0.594                \\
                              & $\mathcal{A}_{inter}$                & 0.788                & 0.690                & 0.599                \\ \hline
\multirow{2}{*}{Ours (AE-U)}  & $\mathcal{A}_{intra}$                & \textbf{0.873}       & 0.743                & \textbf{0.692}       \\
                              & $\mathcal{A}_{inter}$                & \underline{\textbf{0.910}} & \underline{\textbf{0.859}} & \underline{\textbf{0.714}} \\ \hline
\end{tabular}
\end{table}

\section{Conclusion}
In this paper, we propose the Dual-distribution Discrepancy for Anomaly Detection (DDAD), which fully utilizes both known normal and unlabeled images. Two new anomaly scores, intra- and inter-discrepancy, are designed based on DDAD for identifying abnormalities. Experiments on three CXR datasets demonstrate that the proposed DDAD can achieve consistent and significant gains using any reconstruction networks as backbones. The performance reveals an increasing trend with the increasing of AR in the unlabeled dataset, while it also outperforms the reconstruction method when AR is 0. The state-of-the-art performance is achieved by DDAD method built on AE-U. In conclusion, DDAD is the first method that utilizes readily available unlabeled images to improve performace of anomaly detection. We hope this work will inspire researchers to explore anomaly detetcion in a more effective way. 
\subsubsection{Acknowledgement.} 
This work was supported in part by the National Key Research and Development Program of China (grant No. 2018AAA0100400), the National Natural Science Foundation of China (grant No. 62176098, 61872417 and 62061160490), the Natural Science Foundation of Hubei Province of China (grant No. 2019CFA022), and the UGC Grant (grant No. BGF.005.2021).

\bibliographystyle{plain}
\bibliography{paper838}

\end{document}